\documentclass{aidb}
\usepackage{graphicx}
\usepackage{balance}  
\usepackage{booktabs}
\usepackage{url}
\makeatletter
\g@addto@macro{\UrlBreaks}{\UrlOrds}
\makeatother

\graphicspath{{./figures/}}
\usepackage[font=bf,labelfont=bf]{caption}
\usepackage{subcaption}

\usepackage[linesnumbered,ruled,vlined]{algorithm2e}

\usepackage{xspace}

\newcommand{\beforeCaptionSpacing}{\vspace{-0.2cm}}
\newcommand{\afterCaptionSpacing}{\vspace{-0.2cm}}

\usepackage{xcolor}
\usepackage{colortbl}

\begin{document}


\title{The Case for Learned Spatial Indexes}


\numberofauthors{5}

\author{
\alignauthor
Varun Pandey\\
       \affaddr{TUM}\\
       \email{pandey@in.tum.de}
\alignauthor
Alexander van Renen\\
       \affaddr{TUM}\\
       \email{renen@in.tum.de}
\alignauthor
Andreas Kipf\\
       \affaddr{MIT CSAIL}\\
       \email{kipf@mit.edu}
\and
\alignauthor
Ibrahim Sabek\\
       \affaddr{MIT CSAIL}\\
       \email{sabek@mit.edu}
\alignauthor
Jialin Ding\\
       \affaddr{MIT CSAIL}\\
       \email{jialind@mit.edu}
\alignauthor
Alfons Kemper\\
       \affaddr{TUM}\\
       \email{kemper@in.tum.de}
}

\maketitle

\begin{abstract}

Spatial data is ubiquitous. Massive amounts of data are generated every day from billions of GPS-enabled devices such as cell phones, cars, sensors, and various consumer-based applications such as Uber, Tinder, location-tagged posts in Facebook, Twitter, Instagram, etc. This exponential growth in spatial data has led the research community to focus on building systems and applications that can process spatial data efficiently.
In the meantime, recent research has introduced learned index structures.
In this work, we use techniques proposed from a state-of-the art learned multi-dimensional index structure (namely, Flood) and apply them to five classical multi-dimensional indexes to be able to answer spatial range queries. By tuning each partitioning technique for optimal performance, we show that (i) machine learned search within a partition is faster by 11.79\% to 39.51\% than binary search when using filtering on one dimension, (ii) the bottleneck for tree structures is index lookup, which could potentially be improved by linearizing the indexed partitions (iii) filtering on one dimension and refining using machine learned indexes is 1.23x to 1.83x times faster than closest competitor which filters on two dimensions, and (iv) learned indexes can have a significant impact on the performance of low selectivity queries while being less effective under higher selectivities.

\end{abstract}
\section{Introduction}
With the increase in the amount of spatial data available today, the database community has devoted substantial attention to spatial data management. For e.g., NYC Taxi Rides open dataset~\cite{nyctaxidata} consists of pickup and drop-off locations of more than 2.7 billion rides taken in the city since 2009. This represents more than 650,000 taxi rides every day in one of the most densely populated cities in the world, but is only a sample of the location data that is captured by many applications today. Uber, a popular ride hailing service available via a mobile application, operates on a global scale and completed 10 billion rides in 2018~\cite{uber_ten_billion}.
The unprecedented rate of generation of location data has led to a considerable amount of research efforts that have been focused on, systems that scale out~\cite{hadoopgis, distributed_data_store, spatialhadoop, sphinx, stark, locationspark, srx, hadoop_storage, simba, spatialspark, geospark}, databases~\cite{memsql, databases_spatial_comparison, mongodb, oracle, hyperspace}, improving spatial query processing~\cite{hadoop_distance_joins, spatial_join, geojoin, approxjoin, clipped_boxes, paralleljoin, main_memory_joins, two_level_spatial_index, bundled_queries, main_memory_spatial_temporal}, or leveraging modern hardware and compiling techniques~\cite{gpufriendly_1, gpu_friendly_2, spatial_compiler_vision, spatial_compiler_2, spatial_compiler, gpurasterization}, to handle the increasing demands of applications today.

\begin{figure}[t!]
    \centering
    \includegraphics[width=0.8\linewidth]{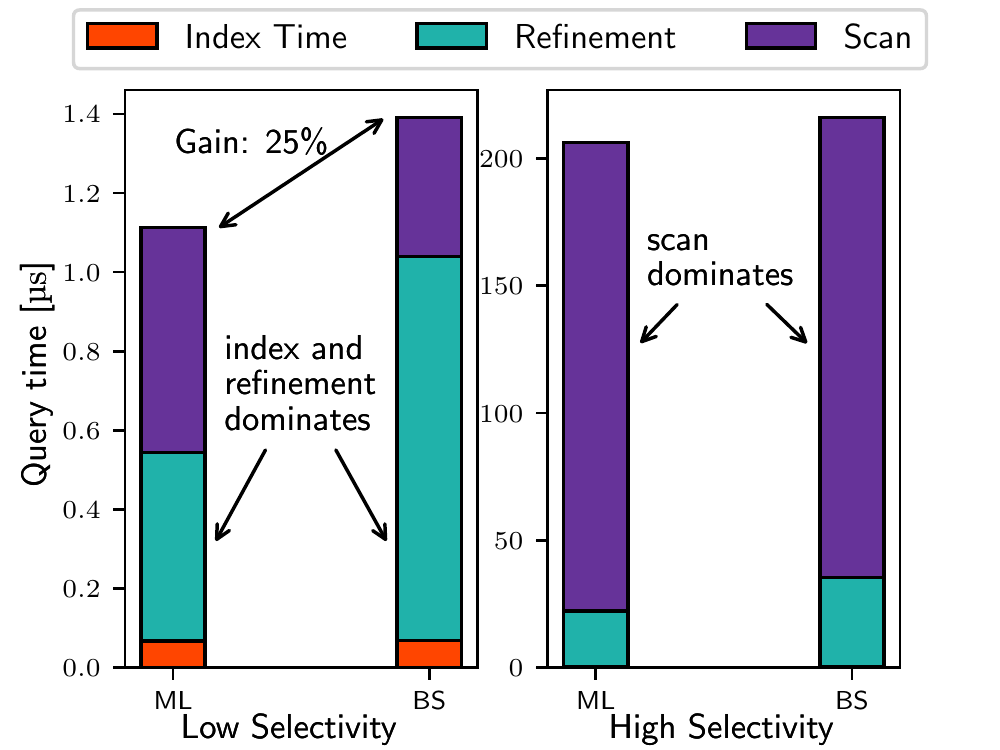}
    \caption{Machine Learning vs. Binary Search. For low selectivity (0.00001\%), the index and refinement phases dominate, while for high selectivity (0.1\%), the scan phase dominates (parameters are tuned to favor Binary Search).}
    \label{fig:mlvsbs}
\end{figure}

\begin{figure*}[htbp!]
	\centering
	\begin{subfigure}{0.18\textwidth}
		\centering
		\scalebox{.24}{\includegraphics{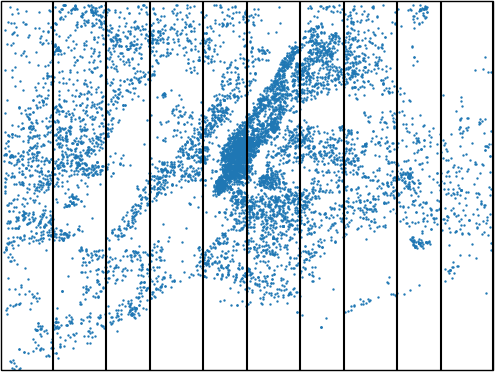}}
		\caption{Fixed grid}
	\end{subfigure}
	\hspace{2mm}
	\begin{subfigure}{0.18\textwidth}
		\centering
		\scalebox{.24}{\includegraphics{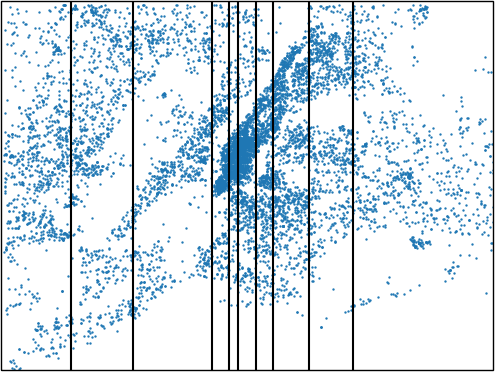}}
		\caption{Adaptive grid}
	\end{subfigure}
	\hspace{2mm}
	\begin{subfigure}{0.18\textwidth}
		\centering
		\scalebox{.24}{\includegraphics{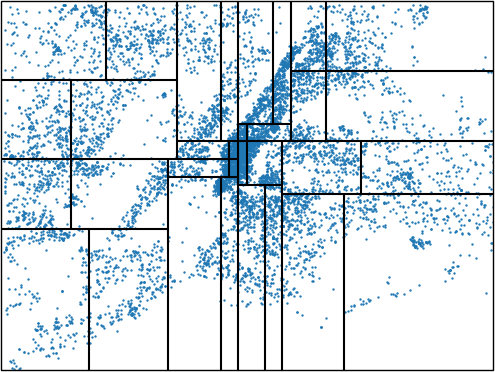}}
		\caption{k-d tree}
	\end{subfigure}
	\hspace{2mm}
	\begin{subfigure}{0.18\textwidth}
		\centering
		\scalebox{.235}{\includegraphics{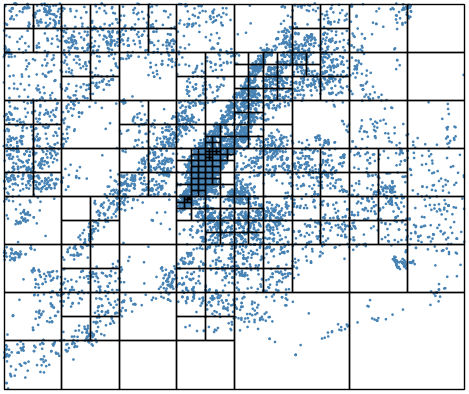}}
		\caption{Quadtree}
	\end{subfigure}
	\hspace{2mm}
	\begin{subfigure}{0.18\textwidth}
		\centering
		\scalebox{.24}{\includegraphics{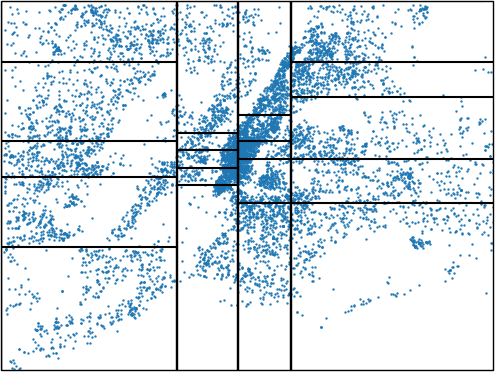}}
		\caption{STRtree}
	\end{subfigure}
	\beforeCaptionSpacing{}
	\caption{\textbf{An illustration of the different partitioning techniques}}
	\label{fig:overview}
	\afterCaptionSpacing{}

\end{figure*}

Recently, Kraska et al.~\cite{rmi} proposed the idea of replacing traditional database indexes with learned models that predict the location of a key in a sorted dataset, and showed that learned models are generally faster than binary search. Kester et al.~\cite{scanorprobe} showed that index scans are preferable over optimized sequential scans in main-memory analytical engines if a query selects a narrow portion of the data. 

In this paper, we build on top of these recent research results, and provide a thorough study for the effect of applying ideas from learned index structures (e.g., Flood~\cite{nathan2020flood}) to classical multi-dimensional indexes. In particular, we focus on five core spatial partitioning techniques, namely Fixed-grid~\cite{fixedgrid}, Adaptive-grid~\cite{gridfile}, Kd-tree~\cite{kdtree}, Quadtree~\cite{quadtree} and STR~\cite{str_packing}. Typically, query processing on top of these partitioning techniques include three phases; index lookup, refinement, and scanning (Details of these phases are in Section~\ref{sec:query_processing}). We propose to replace the typical search techniques used in the refinement phase (e.g., binary search) with learned models (e.g., RadixSpline~\cite{radixspline}).

Interestingly, we found that, by using a learned model as the search technique, we can gain a considerable speedup in the query run-time, especially for low selectivity range queries (Similar to the observation from Kester et al.~\cite{scanorprobe}). Figure~\ref{fig:mlvsbs} shows the average running time of a range query using Adaptive-grid on a Tweets dataset, which consists of 83 million records (Section~\ref{sec:datasets}), with and without learning. It can be seen that for a low selectivity query (which selects 0.00001\% of the data, i.e., 8 records) the index and refinement times dominate the lookup, while for a high selectivity query (which selects 0.1\% of the data, i.e., 83 thousand records) the scan time dominates. Another interesting finding from our study is that 1-dimensional grid partitioning techniques (e.g., Fixed-grid) can benefit from the learned models more than 2-dimensional techniques (e.g., Quadtree). Our study will assist researchers and practitioners in understanding the performance of different spatial indexing techniques when combined with learned models.
\section{Approach}\label{sec:impl}
In this section, we first explain how a range query processing has been implemented. Then, we describe the spatial partitioning techniques that we have implemented in our work.  We conclude the section by describing the search techniques used within the individual partitions.

\subsection{Range Query Processing}\label{sec:query_processing}
A given range query has a lower bound and an upper bound in both dimensions. The task is to materialize all the points that lie within the bounds of the query. Query processing works in three phases:
\begin{itemize}
  \item \textbf{Index Lookup}: In index lookup, we intersect a given range query using the grid directories (or trees) to find the partitions the query intersects with.
  \item \textbf{Refinement}: Once the partitions intersected have been determined from the index lookup phase, we use a search technique (Section~\ref{sec:search}) to find the lower bound of the query on the sorted dimension within the partition.
  There can be various cases on how a query intersects with the partition, and we only consult the search technique when it is actually needed to find the lower bound of the given query on the sorted dimension.. For example, a partition could be fully inside the range query, and in such a case we simply copy all the points in the partition rather than use a search technique.
  \item \textbf{Scan}: Once the lower bound in the sorted dimension has been determined in refinement, we scan the partition to find the qualifying points on both dimensions. We stop as soon as we reach the upper bound of the query on the sorted dimension, or we reach the end of the partition.
\end{itemize}

\subsection{Partitioning Techniques}
Spatially partitioning a dataset into partitions (or cells), such that the objects within the partitions are also close in space, is known as spatial partitioning. Spatial partitioning techniques can be classified into space partitioning techniques (partitions the embedded space) or data partitioning techniques (partitions the data space). In this paper, we employ Fixed-grid~\cite{fixedgrid}, Adaptive-grid~\cite{gridfile}, and Quadtree~\cite{quadtree} as space partitioning techniques; and Sort-Tile-Recursive~\cite{str_packing} and K-d tree~\cite{kdtree} as data partitioning techniques. Figure~\ref{fig:overview} illustrates these techniques on a sample of the Tweets dataset used in our experiments (details are in Section~\ref{sec:datasets}), where sample points and partition boundaries are shown as dots and grid axes respectively.

\subsubsection{Fixed and Adaptive Grid}
The grid (or cell) methods were primarily designed to optimize retrieval of records from disk and generally they share a similar structure. The grid family imposes a d-dimensional grid on the d-attribute space. Every cell in the grid corresponds to one data page (or bucket) and the data points that fall within a particular cell boundary resides in the data page of that cell. Every cell thus has to store a pointer to the data page it indexes. This mapping of grid cells to data pages is known as the grid directory. The Fixed-grid~\cite{fixedgrid} method requires that the grid subdivision lines to be equidistant. The Grid File~\cite{gridfile}, or the Adaptive-grid, on the other hand relaxes this restriction. Since the grid subdivision lines are not equidistant in the case of Grid File, it introduces an auxiliary data structure called linear scales, which are a set of d-dimensional arrays and define the partition boundaries of the d-dimensions. Flood~\cite{nathan2020flood} is a state-of–the–art learned multi-dimensional index for d-dimensional data, which partitions the data using a grid over d-1 dimensions and uses the last dimension as the sort dimension. In our implementation, the grid partitioning techniques use a similar approach where the space is divided in one dimension and the other dimension is used as the sort dimension.

\subsubsection{Quadtree}
Quadtree~\cite{quadtree} along with its many variants is a tree data structure that also partitions the space like the k-d tree. The term quadtree is generally referred to the two-dimensional variant, but the basic idea can easily be generalized to d dimensions. Like the k-d tree, the quadtree decomposes the space using rectilinear hyperplanes. The important distinction is that quadtree is not a binary tree, and the interior nodes in the tree have $2^{d}$ children for d-dimensions. For d = 2, each interior node has four children, each corresponding to a rectangle. 
The search space is recursively decomposed into four quadrants until the number of objects in each quadrant is less than a predefined threshold (usually the page size). Quadtrees are generally not balanced as the tree goes deeper for the areas with higher densities.

\subsubsection{K-d tree}
K-d tree~\cite{kdtree} is a binary search tree that recursively subdivides the space into equal subspaces by means of rectilinear (or iso-oriented) hyperplanes. The subdivision alternates between the k dimensions to be indexed. The splitting hyperplanes at every level are known as the discriminators. For k = 2, for example, the splitting hyperplanes are alternately perpendicular to the x-axis and the y-axis, and are called the x-discriminator and the y-discriminator respectively. The original K-d tree partitioned the \emph{space} into equal partitions, for example if the input space consists of GPS co-ordinate system (-90.0, -180 to 90, 180) the space would be divided into equal halves (-45, -90 to 45, 90). K-d trees are thus not balanced if the data is skewed (most of which might only lie in one partition). K-d tree can be made data-aware by selecting a median point from the data and dividing the data into two halves. This ensures that both partitions in the binary tree are balanced. We have implemented the data-aware k-d tree in our work.

\subsubsection{Sort-Tile-Recursive (STR) packed R-tree}

Sort-Tile-Recursive~\cite{str_packing} is a packing algorithm to fill R-tree~\cite{rtree} and aims to maximize space utilization. The main idea behind STR packing is to tile the data space into $S\times S$ grid. For example, consider the number of points in a dataset to be $P$ and $N$ be the capacity of a node. The data space can then be divided into $S\times S$ grid where $S = \sqrt{P/N}$. The points are first sorted on the x-dimension (in case of rectangles, the x-dimension of the centroid) and then divided into $S$ vertical \emph{slices}. Within each vertical slice, the points are sorted on the y-dimension, and packed into nodes by grouping them into runs of length $N$ thus forming $S$ horizontal slices. The process then continues recursively. Packing the R-tree in this way packs all the nodes completely, except the last node which may have fewer than $N$ elements.

\subsection{Search Within Partition}
\label{sec:search}

The learned index structures require the underlying data to be sorted. In multi-dimensions, there is no inherent sort order over all dimensions. Thus, after partitioning the data, a sort ordering on some dimension is required for the learned indexes to work. To achieve that, within each partition we sort the data using one dimension. Since spatial data consists of two dimensions (in two-dimensional space), either of the two dimensions can be selected as the sort dimension. Once the data within the partition has been sorted, either a learned index or binary search (hereby search technique) can be used on the sorted dimension. In all our experiments, we have sorted on the longitude value of the location.

We use a RadixSpline~\cite{radixspline, radixspline2} over the sorted dimension which consists of two components: 1) a set of spline points, and 2) a radix table to quickly determine the spline points to examine for a lookup key (in our case the dimension over which the data is sorted). At lookup time, first the radix table is consulted to determine the range of spline points to examine. In the next step, these spline points are searched over to determine the spline points surrounding the lookup key. In the last step, linear interpolation is used to predict the position of the lookup key in the sorted array.
Unlike the RMI~\cite{rmi}, the RadixSpline only requires one pass over the data to build the index, while retaining competitive lookup times.
The RadixSpline and the RMI, at the time of writing, only work on integer values, and we adapted the open-source implementation of RadixSpline to work with floating-point values (spatial datasets generally contain floating point values). In our implementation, we have set the spline error to 32 in all experiments.

It is important to make a distinction between how we use RadixSpline and binary search for refinement. In case of binary search, we do a lookup for the lower bound of the query on the sorted dimension. As learned indexes come with an error, usually a local search is done to find the lookup point (in our case the query lower bound). For range scans, as we do, there can be two cases. The first case is that the estimated value from the spline is lower than the actual lower bound on the sorted dimension. In this case, we scan up until we reach the lower bound on the sorted dimension. In the second case, the estimated value is higher than the actual lower bound, hence, we first scan down to the lower bound, materialize all the points in our way until we reach this bound, and after that we scan up until the query upper bound (or the partition end).
In case the estimated value is lower than the upper bound of the query (i.e. the estimated value is within both query bounds), the second case incurs zero cost for local search as we can scan in both directions until we reach the query bounds within the partition.
\section{Evaluation}
\label{sec:evaluation}
All experiments were run single threaded on an Ubuntu 18.04 machine with an Intel Xeon E5-2660 v2 CPU (2.20\,GHz, 10 cores, 3.00\,GHz turbo)\footnote{CPU: \url{https://ark.intel.com/content/www/us/en/ark/products/75272/intel-xeon-processor-e5-2660-v2-25m-cache-2-20-ghz.html}} and 256\,GB DDR3 RAM. We use the \emph{numactl} command to bind the thread and memory to one node to avoid NUMA effects. CPU scaling was also disabled during benchmarking using the \emph{cpupower} command.

\subsection{Datasets}\label{sec:datasets}

\begin{figure}[t]
\centering
  \begin{subfigure}[b]{.29\linewidth}
  \centering
    \resizebox{\textwidth}{!}{\fbox{\includegraphics{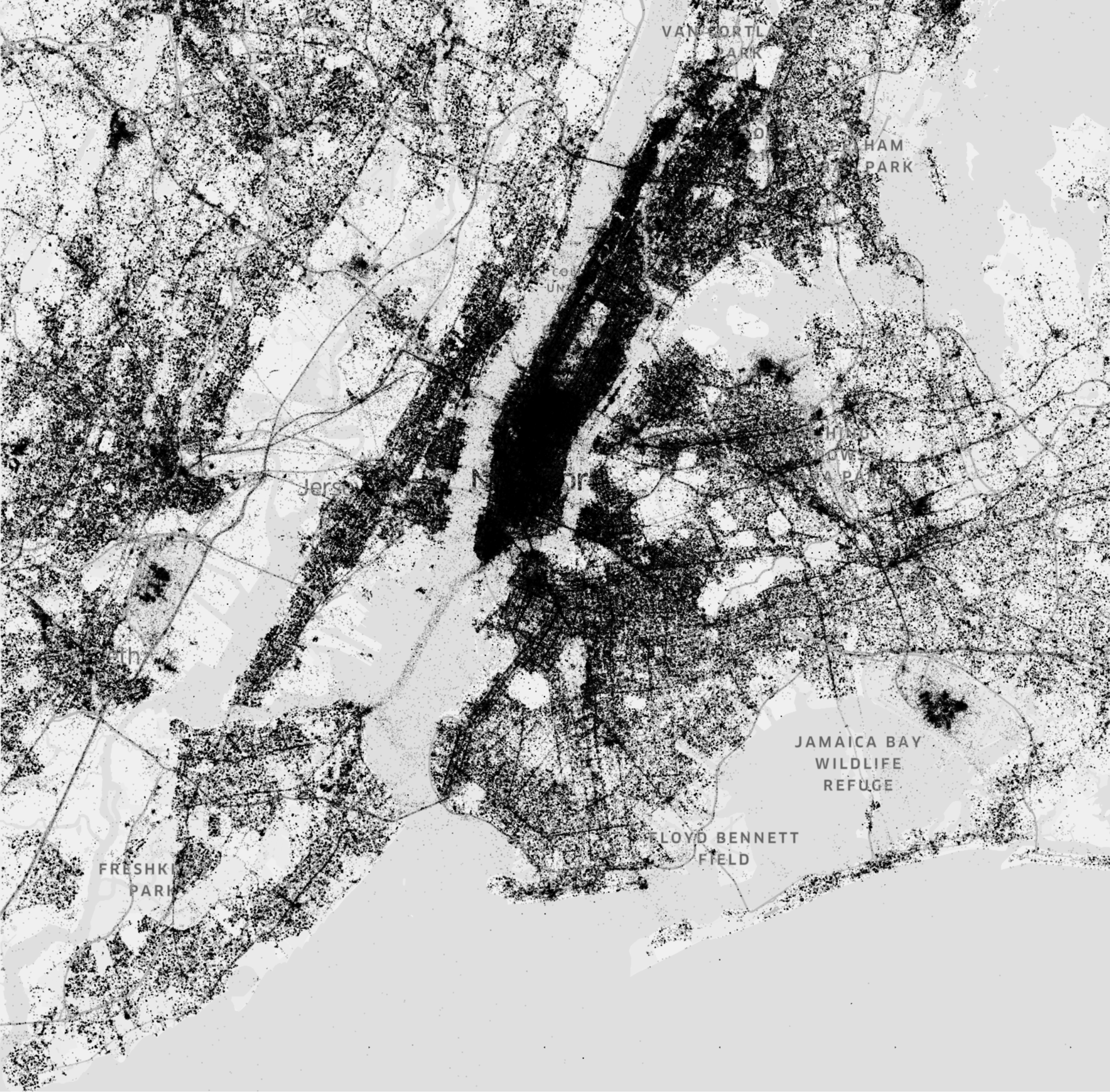}}}
    \caption{Twitter}
    \label{fig:tweets_dataset_bw}
  \end{subfigure}
\hfill
\begin{subfigure}[b]{.29\linewidth}
  \centering
    \resizebox{\textwidth}{!}{\fbox{\includegraphics{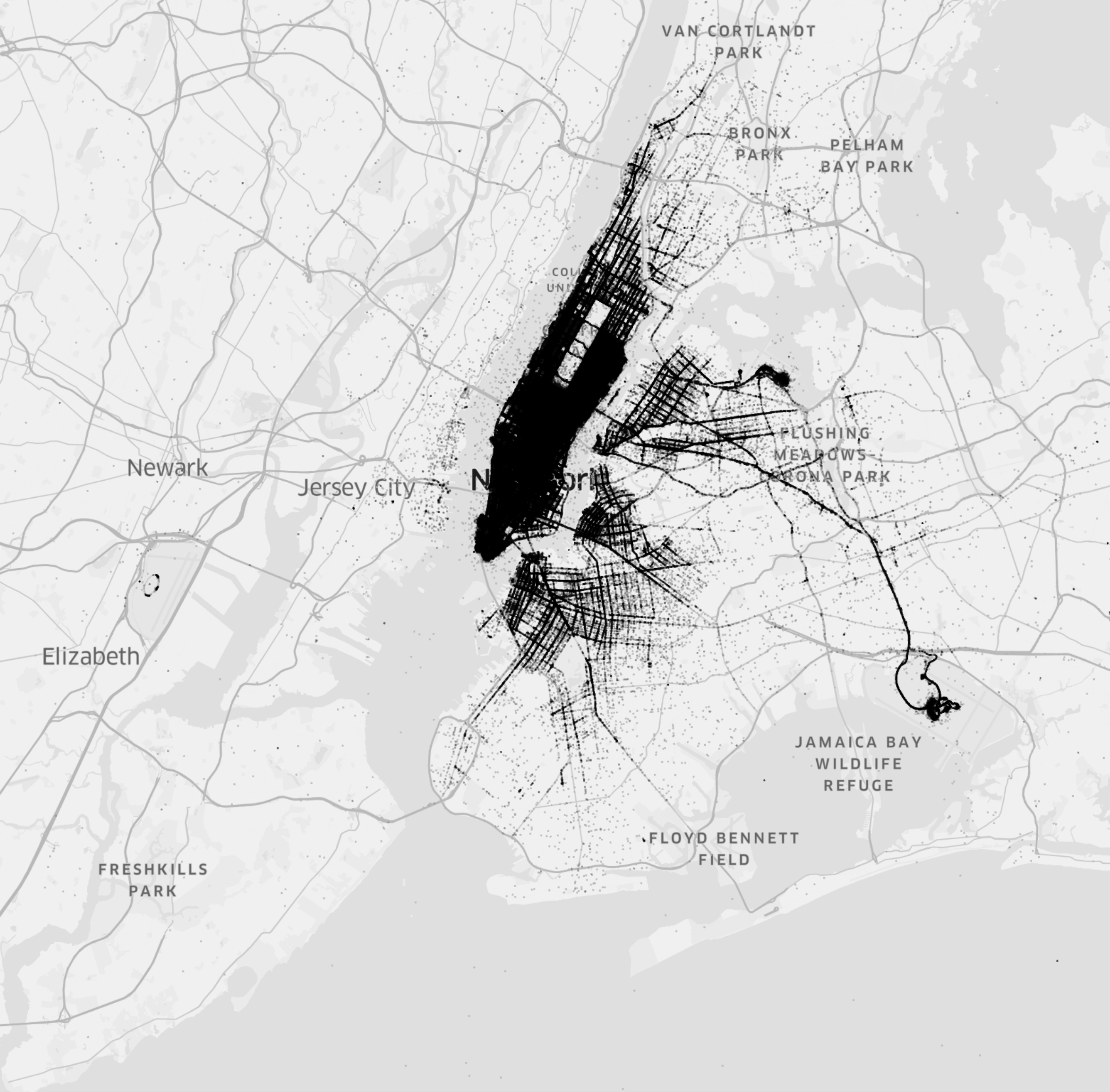}}}
    \caption{Taxi Trips}
    \label{fig:taxi_dataset_bw}
  \end{subfigure}
\hfill
  \begin{subfigure}[b]{.30\linewidth}
  \centering
    \resizebox{\textwidth}{!}{\fbox{\includegraphics{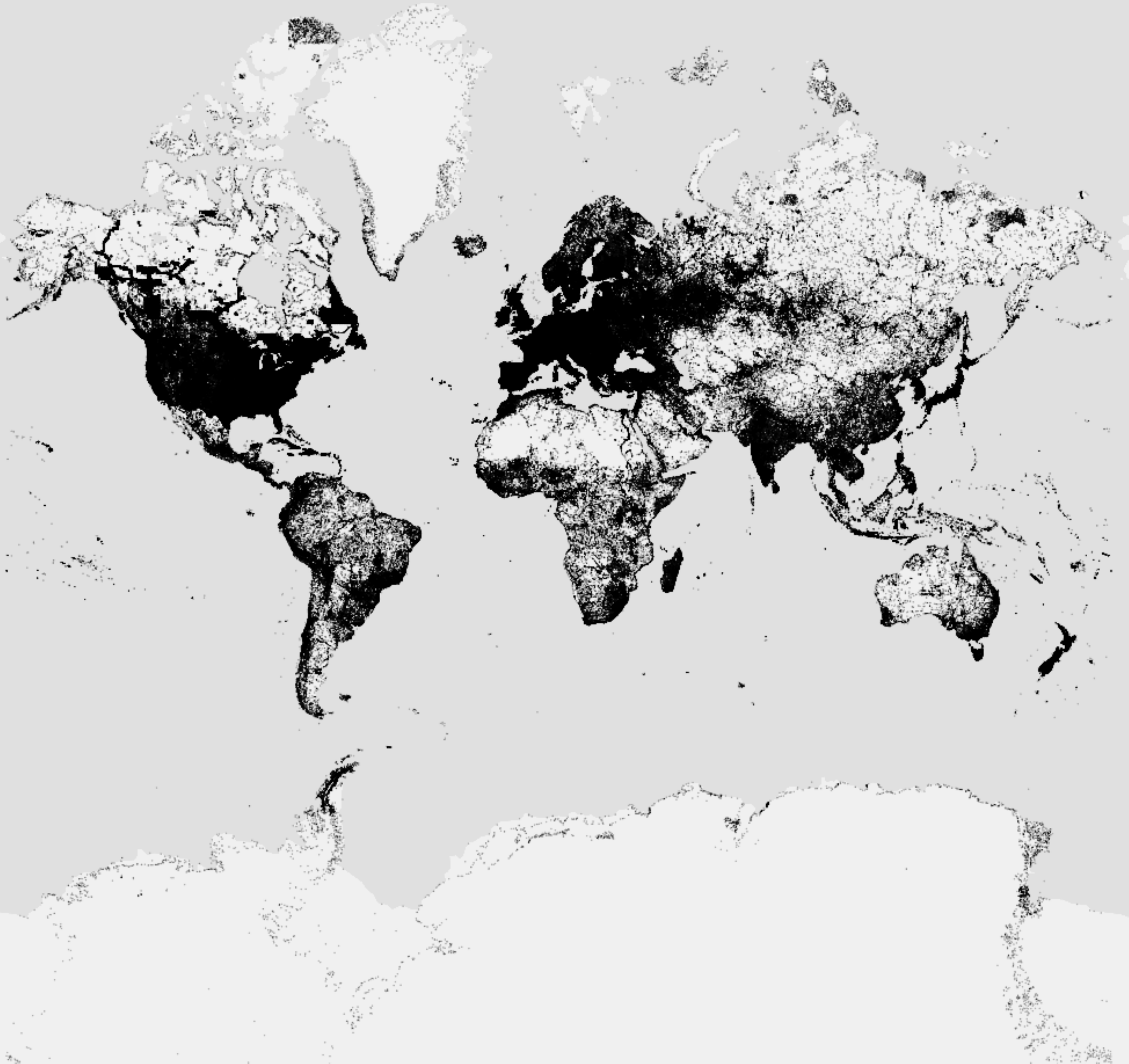}}}
    \caption{OSM}
    \label{fig:osm_dataset_bw}
  \end{subfigure}
\caption{\textbf{Datasets: (a) Tweets are spread across New York, (b) NYC Taxi trips are clustered in central New York, and (c) All Nodes dataset from OSM.}}
\label{fig:datasets_bw}
\end{figure}

For evaluation, we used three datasets, the New York City Taxi Rides dataset~\cite{nyctaxidata} (NYC Taxi Rides), geo-tagged tweets in the New York City area (NYC Tweets), and Open Streets Maps (OSM). NYC Taxi Rides contains 305 million taxi rides from the years 2014 and 2015. NYC Tweets data was collected using Twitter's Developer API~\cite{nyctweets} and contains 83 million tweets. The OSM dataset has been taken from~\cite{howgood} and contains 200M records from the All Nodes (Points) dataset. Figure~\ref{fig:datasets_bw} shows the spatial distribution of the three datasets.
We further generated two types of query workloads for each of the three datasets: skewed queries (which follows the distribution of the underlying data) and uniform queries. For each type of query workload, we generated six different workloads ranging from 0.00001\% to 1.0\% selectivity. For example, in the case of Taxi Rides dataset (305M records), these queries would materialize 30 records to 3 million records. These query workloads consist of one million queries each. To generate skewed queries, we select a record from the data, and expand its boundaries (using a random ratio in both dimensions) until the selectivity requirement of the query is met. For uniform queries, we generated points uniformly in the embedding space of the dataset and expand the boundaries similarly until the selectivity requirement of the query is met. The query selectivity and the type of query are mostly application dependent. For example, consider the application Google Maps, and a user issues a query to find the popular pizzeria near the user. The expected output for this query should be a handful of records, i.e. a low selectivity query (a list of 20-30 restaurants near the user). On the other hand a query on an analytical system, would materialize many more records (e.g. find average cost of all taxi rides originating in Manhattan).

\subsection{Tuning Partitioning Techniques}

\begin{figure*}[htbp]
    \includegraphics[width=\textwidth, height=6.3cm]{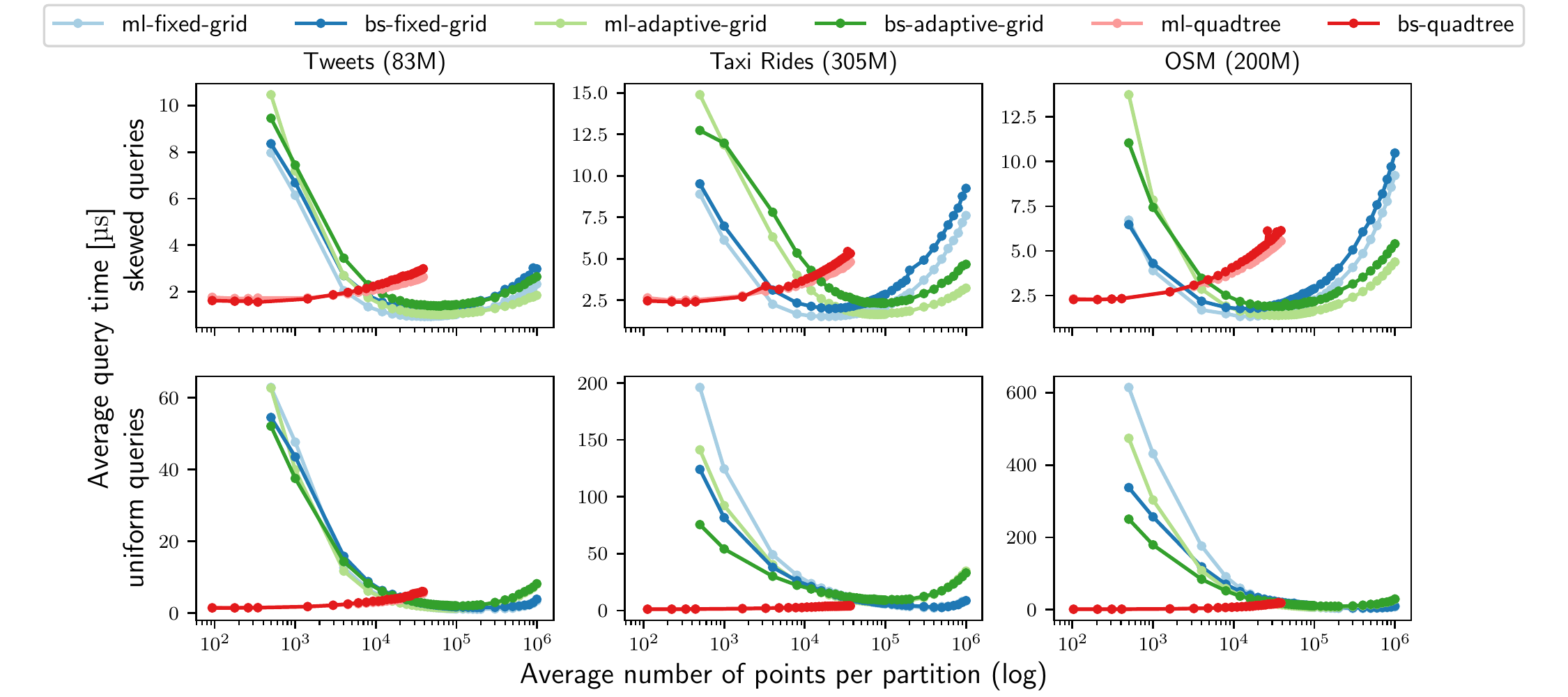}
    \caption{Configuration Experiments - ML vs. BS for low selectivity (0.00001\%).}
    \label{fig:learning_vs_bs_low}
\end{figure*}

\begin{figure}[htbp]
    \centering
    \includegraphics[width=0.8\linewidth]{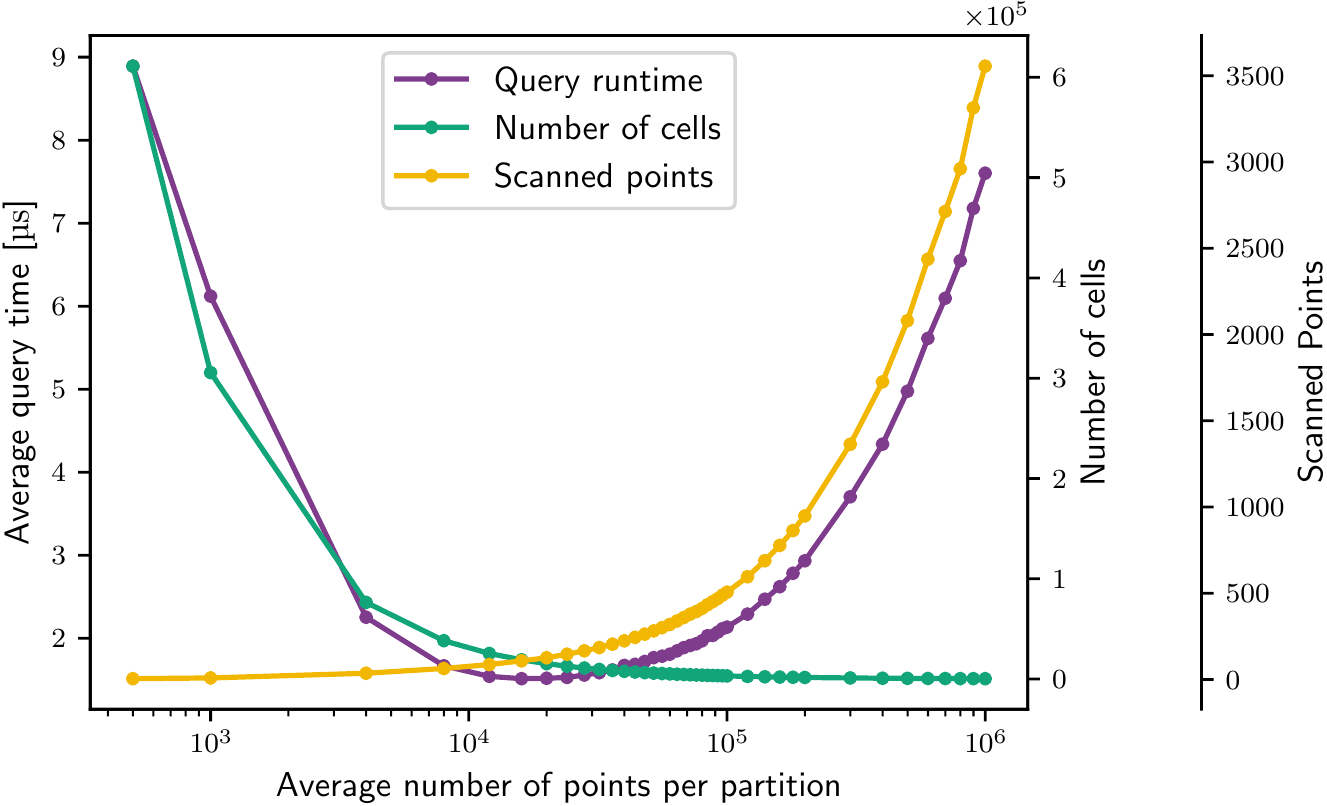}
    \caption{Effect of number of cells and number of points scanned for Fixed-grid on Taxi Trip dataset for skewed queries (0.00001\% selectivity).}
    \label{fig:effect_of_cells}
\end{figure}

\begin{figure}[htbp]
    \centering
    \includegraphics[width=0.85\linewidth]{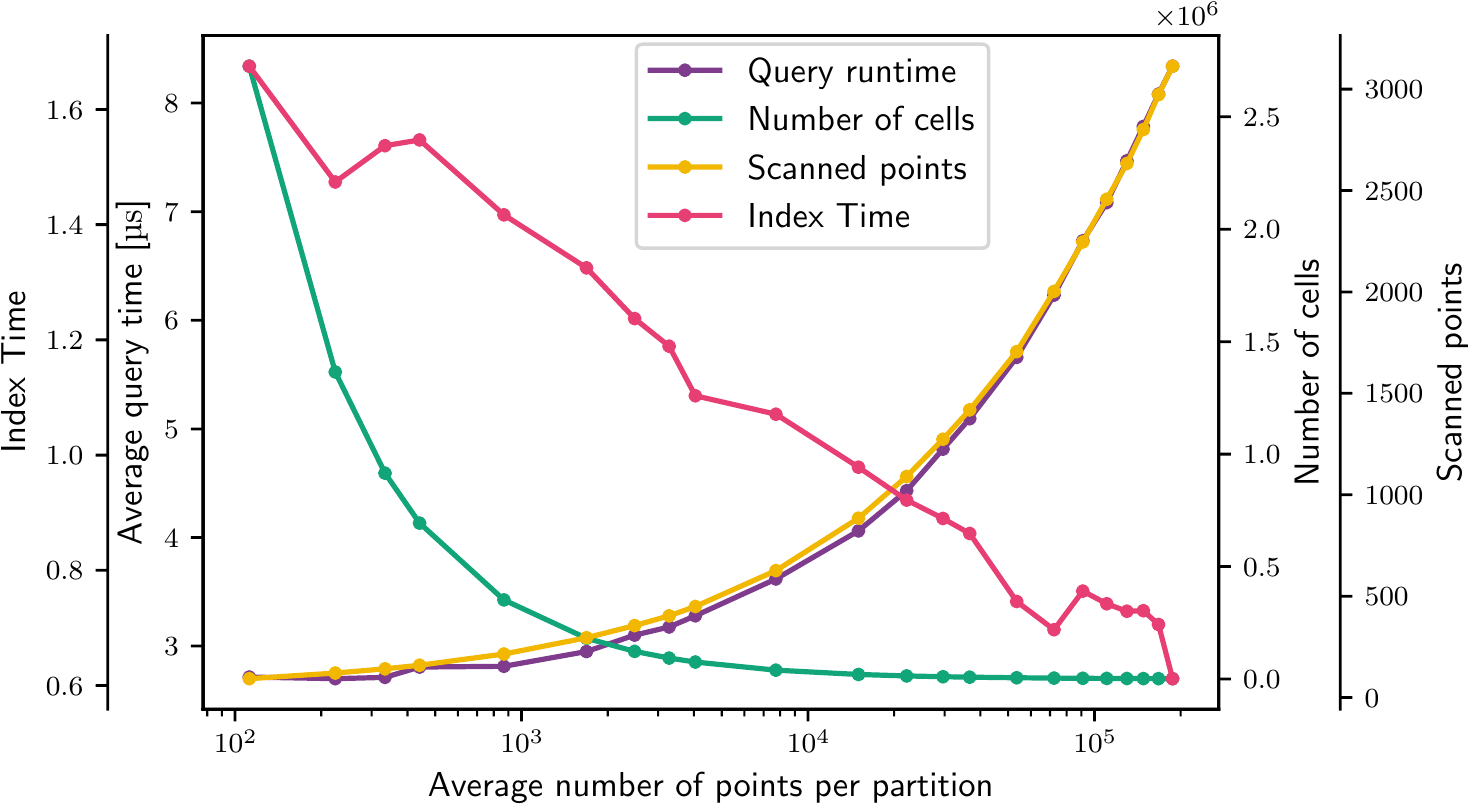}
    \caption{Effect of number of cells and number of points scanned for Quadtree on Taxi Trip dataset for skewed queries (0.00001\% selectivity).}
    \label{fig:effect_of_cells_quadtree}
\end{figure}

Recent work in learned multi-dimensional and spatial indexes have focused on learning from the data and the query workload. The essential idea behind learning from both data and query workload is that a particular usecase can be instance-optimized. To study this effect, we conducted multiple experiments on the three datasets by varying the sizes of the partitions, tuning them on two workloads with different selectivities (to cover a broad spectrum we tune the indexes on queries with low and high selectivity) for both skewed and uniform queries.

Figure~\ref{fig:learning_vs_bs_low} shows the effect of tuning when the indexes are tuned for the lowest selectivity workload for the two query types. It can be seen in the figure that it is essential to tune the grid partitioning techniques for a particular workload. Firstly, they are susceptible to the size of the partition. As the size of the partition increases, we notice an improvement in the performance until a particular partition size is reached which corresponds to the optimal performance. After this point, increasing the size of the partitions only degrades performance. It can be seen that, usually, for grid (single-dimension) partitioning techniques the partition sizes are much larger compared to partitioning techniques which filter on both dimensions (only Quadtree is shown in the figure but the same holds for the other partitioning techniques we have covered in this work, we do not show the other trees because the curve is similar for them). Due to the large partition sizes in grid partitioning techniques, we notice a large increase in performance while using a learned index compared to binary search. This is especially evident for skewed queries (which follow the underlying data distribution). We encountered a speedup from 11.79\% up to 39.51\% compared to binary search. Even when we tuned a learned index to a partition size which corresponds to the optimal performance for binary search, we found that in multiple cases learned index frequently outperformed binary search. Learned indexes do not help much for partitioning techniques which filter on both dimensions, instead the performance of Quadtree (and STRtree) dropped in many cases, see Table~\ref{tab:runtime}. The reason is that the optimal partition sizes for these techniques is very low (less than 1,000 points per partition for most configurations). The refinement cost for learned indexes is an overhead in such cases. K-d tree on the other hand, contains more points per partition (from 1200 to 7400) for the optimal configuration for Taxi Trips and OSM datasets and thus learned indexes perform faster by 2.43\% to 9.17\% than binary search. For Twitter dataset, the optimal configuration contains less than 1200 points per partition, and we observed a similar drop in performance using learned indexes.

Figure~\ref{fig:effect_of_cells} shows the effect of number of cells and number of points that are scanned in each partition on query runtime for Fixed-grid on Taxi Trips dataset for lowest selectivity. As the number of points per partitions increases (i.e. fewer number of partitions), the number of cells decreases. At the same time, the number of points that need to be scanned for the query increases. The point where these curves meet is the optimal configuration for the workload which corresponds to the lowest query runtime. For tree structures, the effect is different. Figure~\ref{fig:effect_of_cells_quadtree} shows that the structures that filter on both dimensions do most of the pruning in the index lookup. The dominating cost in these structures is the number of points scanned within the partition and the query runtime is directly proportional to this number. To minimize the number of points scanned, they do most of the pruning during index lookup which require more partitions (i.e. less number of points per partition), but then they pay more for index lookup.

\subsection{Range Query}
\begin{figure*}[htbp]
    \includegraphics[width=\textwidth]{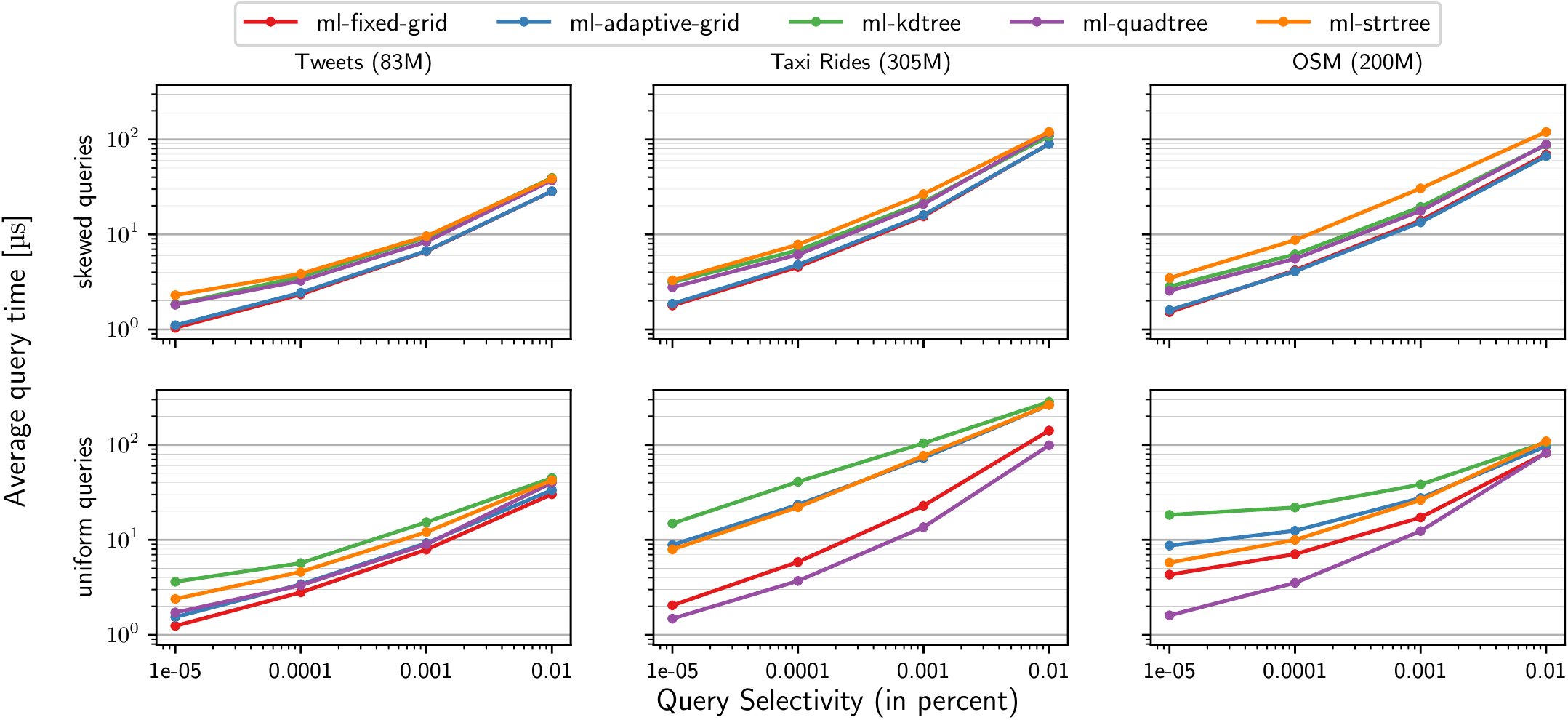}
    \caption{Total query runtime with parameters tuned on selectivity 0.00001\%.}
    \label{fig:runtime_low}
\end{figure*}

Figure~\ref{fig:runtime_low} shows the query runtime for all learned index structures. It can be seen that Fixed-grid along with Adaptive-grid performs (1D schemes) perform the best for all the cases except uniform queries on Taxi and OSM datasets. For skewed queries, Fixed-grid is 1.23$\times$ to 1.83$\times$ faster than the closest competitor, Quadtree (2D), across all datasets and selectivity. The slight difference in performance between Fixed-grid and Adaptive-grid comes from the index lookup. For Adaptive-grid, we use binary search on the linear scales to find the first partition the query intersects with. For Fixed-grid, the index lookup is almost negligible as only an offset computation is needed to find first intersecting partition. It can also be seen in the figure that the Quadtree is significantly better for uniform queries in case of Taxi Rides dataset (1.37$\times$) and OSM dataset (2.68$\times$) than the closest competitor Fixed-grid. There are two reasons for this, firstly the Quadtree intersects with fewer number of partitions than the other index structures, see Table~\ref{tab:stats}. Secondly, for uniform queries, the Quadtree is more likely to traverse the sparse and low-depth region of the index. This is in conformance with an earlier research~\cite{quadtree_uniform}, where the authors report similar findings while comparing the Quadtree to the R*-tree and the Pyramid-Technique.

\begin{table*}
\scriptsize
\begin{tabular}{lrrrrrrrrrrrrrrrrrrrr}
\toprule
{} & \multicolumn{6}{c}{Taxi Trips (Skewed Queries)} & \multicolumn{6}{c}{Taxi Trips (Uniform Queries)} \\
\midrule
{} & \multicolumn{2}{c}{Fixed} & \multicolumn{2}{c}{Adaptive} & \multicolumn{2}{c}{Quadtree} & \multicolumn{2}{c}{Fixed} & \multicolumn{2}{c}{Adaptive} & \multicolumn{2}{c}{Quadtree} \\
\midrule
{Selectivity (\%)}       &  ML       &    BS      &  ML       &  BS       &  ML        &  BS       &  ML        &  BS       & ML       &  BS       &    ML       &    BS   \\
\midrule
0.00001             &  1.78    &   2.35    &  1.86    &  2.40    &  2.77     &  2.51    &  2.02     &  2.58    & 81.4    &  10.54   &  1.48      &  1.31  \\
0.0001              &  4.54    &   5.82    &  4.67    &  6.12    &  6.12     &  5.82    &  5.85     &  6.91    & 228.1   &  27.69   &  3.69      &  3.42   \\
0.001               &  14.97   &   18.83   &  15.32   &  19.49   &  20.84    &  19.47   &  22.87    &  24.34   & 708.8   &  87.49   &  13.59     &  12.98  \\
0.01                &  90.13   &   97.04   &  89.48   &  95.96   &  117.01   &  104.37  &  141.24   &  151.47  & 2634.4  &  309.62  &  98.85     &  112.77 \\
0.1                 &  678.12  &   698.39  &  675.14  &  696.49  &  922.67   &  793.96  &  988.35   &  922.96  & 9609.9  &  1174.79 &  891.24    &  1101.95 \\
1.0                 &  8333.94 &   8408.15 &  8301.56 &  8399.69 &  10678.04 &  9512.29 &  8843.71  &  8753.68 & 8574.84 &  8836.28 &  10647.97  &  12377.14  \\
\bottomrule
\end{tabular}
\centering
\caption{Total query runtime (in microseconds) for both RadixSpline (ML) and binary search (BS) for Taxi Rides dataset on skewed and uniform query workloads (parameters are tuned for selectivity 0.00001\%).}
\label{tab:runtime}
\end{table*}

\begin{table}
\footnotesize
\begin{tabular}{lrrrr}
\toprule
{} & \multicolumn{2}{c}{Taxi Rides} & \multicolumn{2}{c}{OSM} \\
\midrule
{Partitioning}     &  Skewed   &   Uniform  &   Skewed  &  Uniform  \\
\midrule
Fixed             &  1.97      &   7.98     &   1.72    &  23.73    \\
Adaptive          &  1.74      &   31.57    &   1.51    &  24.80    \\
k-d tree          &  1.70      &   21.62    &   1.56    &  30.95    \\
Quadtree          &  1.79      &   2.12     &   1.37    &  7.96     \\
STR               &  2.60      &   47.03    &   1.90    &  11.05    \\
\bottomrule
\end{tabular}
\centering
\caption{Average number of partitions intersected for each partitioning scheme for selectivity 0.00001\% on Taxi Rides and OSM datasets.}
\label{tab:stats}
\end{table}

\subsection{Indexing Costs}

\begin{figure}[htbp]
    \centering
    \includegraphics[height=5.0cm]{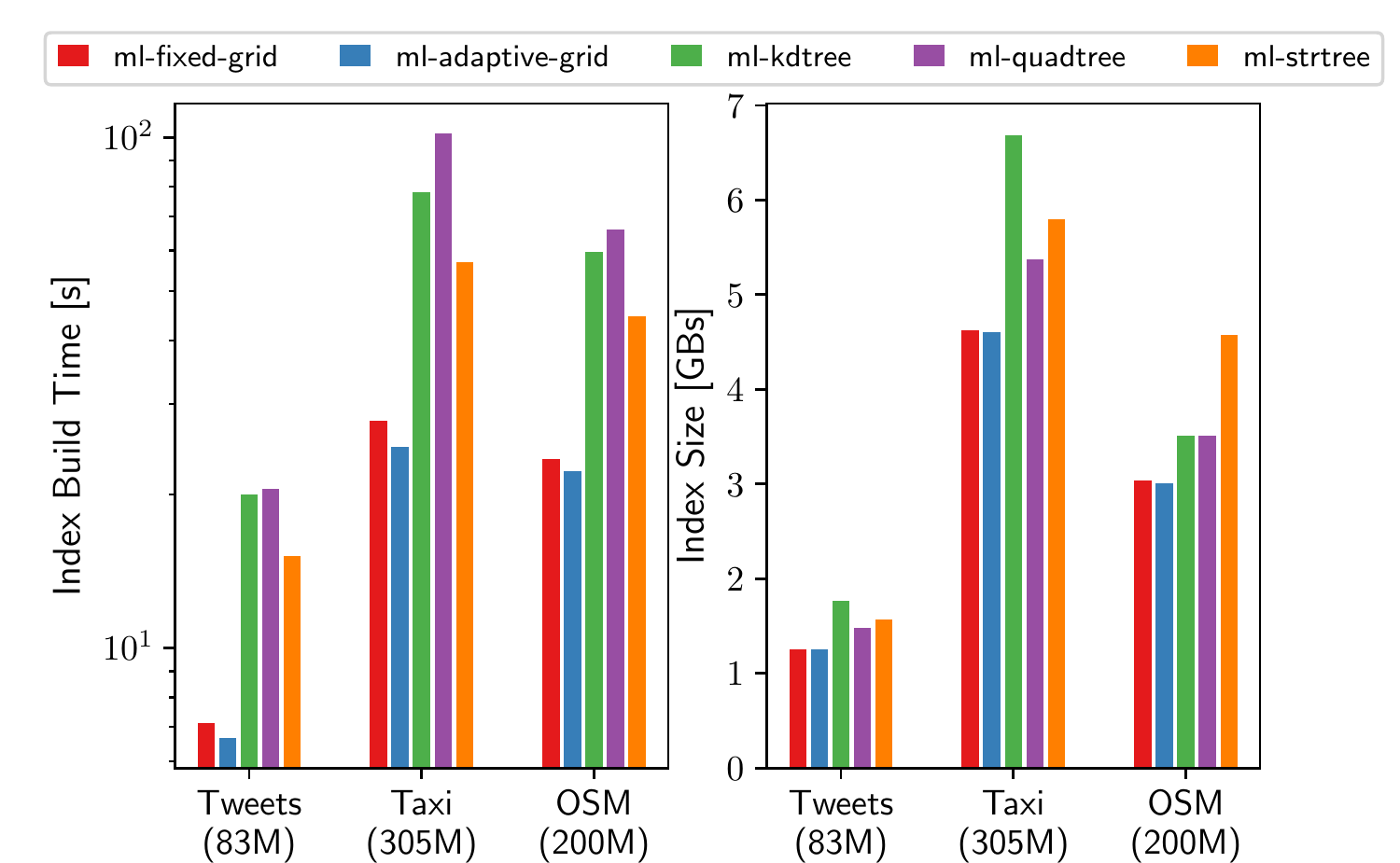}
    \caption{Index build times and sizes for the three datasets.}
    \label{fig:indexing_costs}
\end{figure}

Figure~\ref{fig:indexing_costs} shows that Fixed-grid and Adaptive-grid are faster to build than the tree based indexes. Fixed-grid is 2.11$\times$, 2.05$\times$, and 1.90$\times$ faster to build than closest competitor STRtree. Quadtree is the slowest to build because it generates a large number of cells for optimal configuration. Not all partitions in Quadtree contain an equal number of points as it divides space rather than data, thus leading to an imbalanced number of points per partition. Fixed-grid and Adaptive grid do not generate large number of partitions, as the partitions are quite large for optimal configuration. They are lower in size for similar reasons. The index size in Figure~\ref{fig:indexing_costs} also includes the size of data being indexed.

\section{Related Work}
\label{sec:relatedwork}
Recent work by Kraska et al.~\cite{rmi} proposed the idea of replacing traditional database indexes with learned models that predict the location of one-dimensional keys in a dataset. 
Since then, there has been a corpus of work on extending the ideas of the learned index to spatial and multi-dimensional data.
Flood~\cite{nathan2020flood} is an in-memory read-optimized multi-dimensional index that organizes the physical layout of $d$-dimensional data by dividing each dimension into some number of partitions, which forms a grid over $d$-dimensional space and adapts to the data and query workload.
Learning has also been applied to the challenge of reducing I/O cost for disk-based multi-dimensional indexes. Qd-tree~\cite{qdtree-sigmod} uses reinforcement learning to construct a partitioning strategy that minimizes the number of disk-based blocks accessed by a query. LISA~\cite{lisa-sigmod} is a disk-based learned spatial index that achieves low storage consumption and I/O cost while supporting range queries, nearest neighbor queries, and insertions and deletions.
The ZM-index~\cite{zm-index} combines the standard Z-order space-filling curve with the RMI from~\cite{rmi} by mapping multi-dimensional values into a single-dimensional space, which is then learnable using models. The ML-index~\cite{ml-index} combines the ideas of iDistance~\cite{idistance} and the RMI to support range and KNN queries.

\section{Conclusions and Future Work}
\label{sec:conclusion}

In this work, we implemented techniques proposed in a state-of-the-art multi-dimensional index, namely, Flood~\cite{nathan2020flood}, which indexes points using a variant of the Grid-file and applied them to five classical spatial indexes. We have shown that replacing binary search with learned indexes within each partition can improve overall query runtime by 11.79\% to 39.51\%. As expected, the effect of using a more efficient search within a partition is more pronounced for queries with low selectivity. With increasing selectivity, the effect of a fast search diminishes.
Likewise, the effect of using a learned index is larger for (1D) grid partitioning techniques (e.g., Fixed-grid) than for (2D) tree structures (e.g., Quadtree).
The reason is that the partitions (cells) are less representative of the points they contain in the 1D case than in the 2D case.
Hence, 1D partitioning requires more refinement within each cell.

In contrary, finding the qualifying partitions is more efficient with 1D than with 2D partitioning, thus contributing to lower overall query runtime (1.23x to 1.83x times faster).
Currently, we are using textbook implementations for Quadtree and K-d tree.
Future work could study replacing these tree structures with learned counterparts.
For example, we could linearize Quadtree cells (e.g., using a Hilbert or Z-order curve) and store the resulting cell identifiers in a learned index.

So far we have only studied the case where indexes and data fit into RAM.
For on-disk use cases, performance will likely be dominated by I/O and the search within partitions will be of less importance.
We expect partition sizes to be performance-optimal when aligned with the physical page size.
To reduce I/O, it will be crucial for partitions to not contain any unnecessary points.
Hence, we expect 2D partitioning to be the method of choice in this case.
We refer to LISA~\cite{lisa-sigmod} for further discussions on this topic.

\balance
\clearpage
\bibliographystyle{abbrv}
\bibliography{learnedspatial}

\end{document}